# From Relational and Property Graph Data to Large Language Models

Malcolm Crowe
University of the West of Scotland (retired)
Paisley, UK
e-mail: Malcolm.Crowe@uws.ac.uk

Fritz Laux
Reutlingen University (retired)
Reutlingen, Germany
e-mail: Friedrich.Laux@reutlingen.de

***Abstract*** **— A data management server that combines the relational data model found in traditional databases with the graphical models being developed by newer data products can provide its knowledge model to generative tools for large language models. Both models have been reconciled at the implementation level by replacing foreign keys by reference values. This boosts performance and allows effectively following links instead of lookup joins. The reference value will never change as it uses the record address in the log file. This paper reports on progress in providing such a synthesis and includes an example of a knowledge base that is triple-based and would also be supported in such a system.**

## I. INTRODUCTION

It has been established by experiment [1] that the best way of prompting Artificial Intelligence (AI) generative models is by giving them the vocabulary appropriate to a problem scenario, and the best way of communicating it is as a knowledge graph using serialization format, such as JavaScript Object Notation (JSON). Unfortunately, constructing any sort of knowledge graph for a real problem area is itself a complex and time-consuming task. A good way is to begin with whatever structured data is available. If it is already in a database, get the database to generate JSON from it. There are many products to assist in the process, but they all seem to have complex rules. They should be made easier to use, and this presentation aims to walk through a realistic case study.

This paper is part of an open-source research project [2] to support the approaches of both ISO/IEC 9075 Database Language SQL (Structured Query Language) and ISO/IEC 39075 Database Language GQL (Graph Query Language) in a single stand-alone database server. Section II outlines the current state of the art. The key contribution of this paper is outlined in Section III, and Section IV describes a case study showing some of the new approaches to data thus enabled. Some important highlights of the implementation are detailed in Section V. Section VI provides some performance measurements, and Section VII details some conclusions and planned future work. This will include contributions towards the emerging technologies for using large language models.

## II. RELATED WORK

SQL/PGQ [3], [4], [5] and Neo4j [6] offer standalone server-based implementations of graph databases, but neither was designed to provide the advanced semantics envisaged in the GQL standard, such as MATCH over repeating graph patterns and horizontal aggregations. The review of the field offered by Kondylakis et al. [7] identified current challenges in the field: (a) Support for schema constraints remains limited and fragmented in Graph Database Management Systems (GDBMSs) and applications, hindering data integration, query optimization and Machine Learning (ML) feature extraction; (b) Data integration is particularly difficult with Resource Description Framework (RDF) graphs and requires either mapping both models to a common unifying representation, while direct transformations between the two have difficulty in ensuring that RDF triples are accurately mapped to the nodes, edges, and properties of property graphs and vice versa; (c) Normalization and standard forms are fundamentally under-researched (but see [8]); (d) GQL was designed to allow massively parallel execution, but graph query answering is relatively slow in current implementations. It has been observed [9], that the only mechanism in SQL for linking data is foreign keys, but SQL does not make foreign keys usable in queries.

Meantime, almost all graphical databases are built by chaining together transformations provided by different products. The standardization process continues however, and new editions of the international standards are expected in 2027.

## III. THE RESEARCH CONTRIBUTION

As mentioned above, SQL uses key values to enable tabular data to be joined, while edges in GQL join individual nodes. In previous work [10], we have shown that for record-based data the concept of reference values can support both approaches, so that a unified approach to linked data management should use structural reference values instead of keys. A bonus is that a relational system holding such data can then support a range of alternative graph models. That work outlined an ongoing implementation of a SQL server that supports GQL syntax, graph insertion, and graph queries. The main contribution of our paper is to explain the implementation of the hybrid graph-relational database Pyrrho V8. Some details of this design are summarized in Section IV below.

In a record-based Database Management System (DBMS), an edge type record contains one or more references, e.g., (1 reference) a record might identify a company (e.g., in a Person record, WorksIn: Cisco), (2 references) a record of employment of a person in a company (a WorksIn record, Person: Joe, Company: Cisco), and so on. The edge type can contain non-reference properties (e.g., dates). Figure 1 shows a simple example.

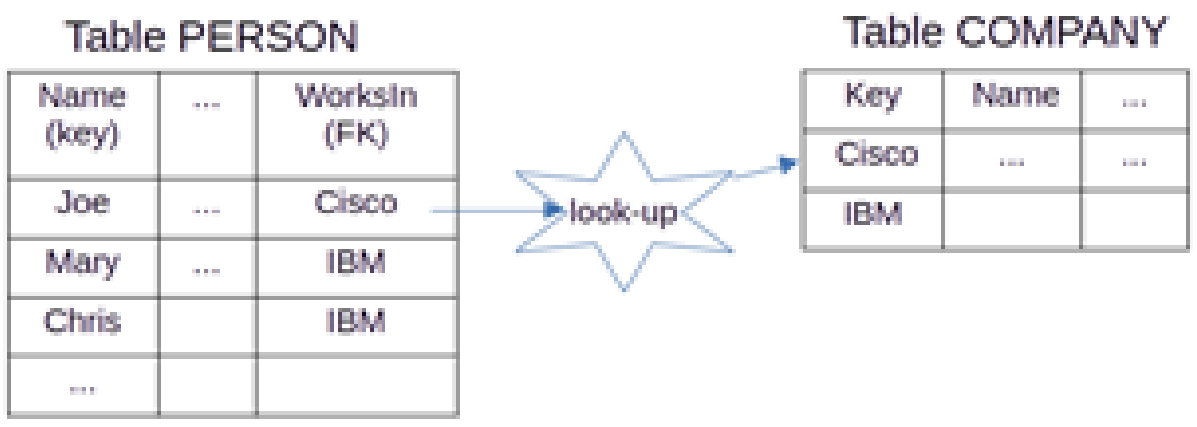

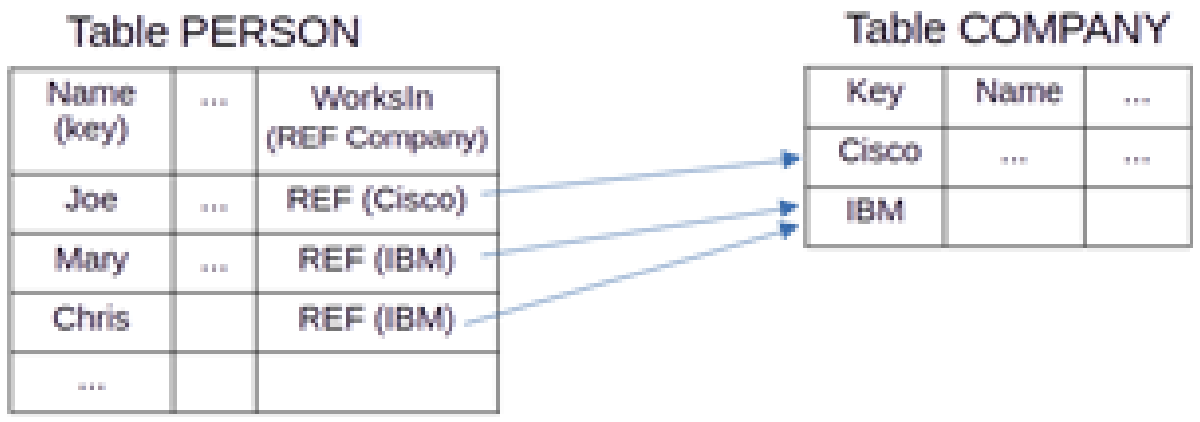

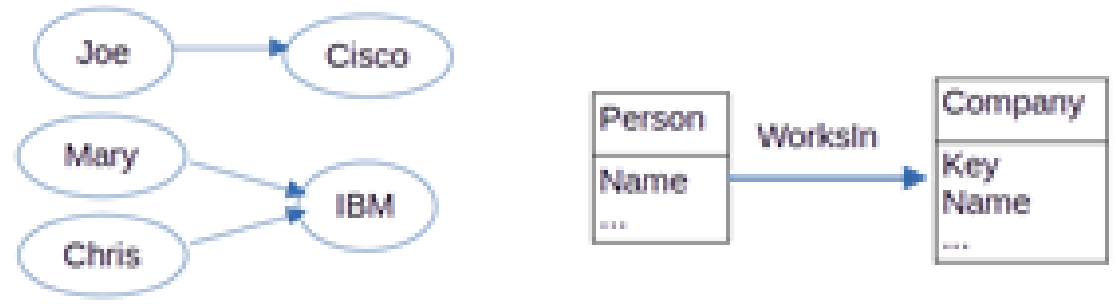

Figure 1. REF and foreign key values.

In the (Person worksIn company) example, the question is – what sort of record is this? It could be an entry in a list of contacts or job applications (Fred, current employer Cisco, born in Scotland..). Do these values identify records in tables of persons, companies, countries? If no, then these properties are not actual references. If yes, then the reference identifies a particular (current, etc.) company for the person, etc.

Given a database with such records, we can define graph models by ignoring some node and edge types, ignoring particular properties, and/or identifying two references `FROM` and `TO` in the edge types we retain. In graph models, we use labels for different sorts of information, typically preceded by a colon, and property lists using braces, such as `{product:'Fuel', supplier:'Exxon'}`.

There are numerous graph database products that allow the definition of node and edge types: several products support graph construction using "Ascii-art", graph patterns where nodes are indicated by parentheses and edges by square brackets: e.g.,

```
INSERT (Joe)-[WorksIn]->(Cisco)
```

The goal of this research is to handle all three cases of Figure 1 using reference columns in relational tables. Section V below outlines how a foreign key as in Figure 1(a) can be implemented using reference columns. It is fairly clear that an edge type as Figure 1(c) implies references to the source and destination nodes. In the next section we explore the third case, corresponding to Figure 1(b), where the reference values are bound during insertion of a graph pattern.

## IV. CASE STUDY [11]

One way of modeling data is design by example: and constructing graph patterns for a new database works this way. Let us look at a complex data model, taking the ACM

**AR. Architecture and Organization (0 Core-Tier1 hours, 16 Core-Tier2 hours)**

| | Core-Tier1 hours | Core-Tier2 Hours | Includes Elective |
|---|---|---|---|
| AR/Digital Logic and Digital Systems | | 3 | N |
| AR/Machine Level Representation of Data | | 3 | N |
| AR/Assembly Level Machine Organization | | 6 | N |
| AR/Memory System Organization and Architecture | | 3 | N |
| AR/Interfacing and Communication | | 1 | N |
| AR/Functional Organization | | | Y |
| AR/Multiprocessing and Alternative Architectures | | | Y |
| AR/Performance Enhancements | | | Y |

**AR/Digital Logic and Digital Systems**

*[3 Core-Tier2 hours]*

***Topics:***

- Overview and history of computer architecture
- Combinational vs. sequential logic/Field programmable gate arrays as a fundamental combinational + sequential logic building block
- Multiple representations/layers of interpretation (hardware is just another layer)
- Computer-aided design tools that process hardware and architectural representations
- Register transfer notation/Hardware Description Language (Verilog/VHDL)
- Physical constraints (gate delays, fan-in, fan-out, energy/power)

***Learning outcomes:***

1. Describe the progression of computer technology components from vacuum tubes to VLSI, from mainframe computer architectures to the organization of warehouse-scale computers. [Familiarity]
2. Comprehend the trend of modern computer architectures towards multi-core and that parallelism is inherent in all hardware systems. [Familiarity]

Figure 2. An excerpt from CS2013 document [3].

CS2013 syllabus report [11] as an example. It has 18 Knowledge Areas with dozens of subject areas (called Knowledge Units): each unit has Topics and Learning Outcomes, categorized as Core-Tier1, Core-Tier2, and Elective. Let us assume we can add a number of similar items by nesting in curly brackets. We suddenly find we are quite naturally getting a new sort of model, that is neither relational nor the usual kind of graph.

However, the nature of the data as records is quite clear: there can be a new or modified knowledge area, a new knowledge unit, new or modified level measures or degree programs, or new topics or learning outcomes in a given knowledge unit.

The university probably wants to arrange these into degree programs with modules and years of study. How to get started? We can simply start adding the data we have. If we start with Appendix A of the report, we can simply introduce some bindings for the Categories and Learning Levels and just start pasting in some of the details we see. As we do this, our DBMS can infer the node and edge types we are defining. As above, property lists are enclosed in braces, and nodes are enclosed in parentheses:

```
Insert (c1:Category{name:'Core-Tier1'}),
(c2:Category{name:'Core-Tier2'}),
(el:Category{name:'Elective'}),
(fa:Level{name:'Familiarity'}),
("as":Level{name:'Assessment'}),
(us:Level{name:'Usage'}),
(al:KnowledgeArea{name:'AL'})<-{
(:KnowledgeUnit{name:'AL/Basic Analysis'})<-{
(:Topic{detail:' Differences among best, expected, and
worst case behaviors of an algorithm',category:c1}),
(:Topic{detail:' Asymptotic analysis of upper and
expected complexity bounds',category:c1}),
(:Topic{detail:' Big O notation: formal
definition',category:c1}),
(:Topic{detail:' Complexity classes, such as constant,
logarithmic, linear, quadratic, and
exponential',category:c1}),
(:Topic{detail:' Empirical measurements of
performance',category:c1}),
(:Topic{detail:' Time and space trade-offs in
algorithms',category:c1}),
(:Topic{detail:' Big O notation: use',category:c2}),
(:Topic{detail:' Little o, big omega and big theta
notation',category:c2}),
(:Topic{detail:' Recurrence relations',category:c2}),
(:Topic{detail:' Analysis of iterative and recursive
algorithms',category:c2}),
(:Topic{detail:' Some version of a Master
Theorem',category:c2}),
(:LearningOutcome{detail:' Explain what is meant by
“best”, “expected”, and “worst” case behavior of an
algorithm.',category:c1,Level:fa}),
(:LearningOutcome{detail:' In the context of specific
algorithms, identify the characteristics of data and/or
other conditions or assumptions that lead to different
behaviors.',category:c1,Level:"as"}),
(:LearningOutcome{detail:' Determine informally the
time and space complexity of simple algorithms.',
category:c1,Level:us})
```

There are hundreds of pages, but it is possible to stop and examine the model once some data has been loaded: simply close both braces: }} .

So far, the DBMS transaction log is as shown in Table I.

There is a single transaction here that has created 6 node types, 4 reference types, 8 named properties, and 22 records. Of course there is much more of CS2013 to insert.

To resume the load, (a) use Match to reestablish the bindings for c1, c2, el, fa, as, us, and al, (b) add a binding for the knowledge unit currently worked on, which was ‘AL/Basic Analysis', and (c) resume the insert operation. If it was switched off, the database reloads the work done already:

```
Match (c1:Category{name:'Core-Tier1'}),
(c2:Category{name:'Core-Tier2'}),
(el:Category{name:'Elective'}),
(fa:Level{name:'Familiarity'}),
("as":Level{name:'Assessment'}),
(us:Level{name:'Usage'}),
(alba:KnowledgeUnit{name:'AL/Basic Analysis'})
Insert (alba)<-{
(:LearningOutcome{detail:' State the formal definition
of big O.',category:c1,Level:fa}),
(:LearningOutcome{detail:' List and contrast standard
complexity classes.',category:c1,Level:fa})}
```

This time we are only one brace deep. Let us look at the data model we have so far:

```
KnowledgeArea: (name) <- KnowledgeUnit: (name) <-
{ Topic: (detail,category),
LearningOutcome: (detail, category, level) }
```

We can already see that some data relationships are being handled by simple properties, and others, such as the reference of a KnowledgeUnit (knowledge unit) to its knowledge area, by simple edges. Figure 3 attempts to show this structure.

TABLE I. TRANSACTION LOG FOR CS2013

| 5 | PTransaction for 44 Role=-502 User=-501 Time=14/04/2026 09:39:04 |
|---|---|
| 23 | PNodeType CATEGORY NODETYPE |
| 49 | PNodeType LEVEL NODETYPE |
| 72 | PNodeType KNOWLEDGEAREA NODETYPE |
| 103 | PNodeType KNOWLEDGEUNIT NODETYPE |
| 134 | PNodeType TOPIC NODETYPE |
| 157 | Domain REF 23 |
| 171 | Domain REF 103 |
| 185 | PNodeType LEARNINGOUTCOME NODETYPE |
| 218 | Domain REF 49 |
| 232 | Domain REF 72 |
| 246 | PColumn3 NAME for 23(0)[ CHAR] NOT NULL |
| 269 | Record 269[23] 246=Core-Tier1[CHAR] |
| 292 | Record 292[23] 246=Core-Tier2[CHAR] |
| 315 | Record 315[23] 246=Elective[CHAR] |
| 336 | PColumn3 NAME for 49(0)[ CHAR] NOT NULL |
| 359 | Record 359[49] 336=Familiarity[CHAR] |
| 383 | Record 383[49] 336=Assessment[CHAR] |
| 406 | Record 406[49] 336=Usage[CHAR] |
| 424 | PColumn3 NAME for 72(0)[ CHAR] NOT NULL |
| 447 | Record 447[72] 424=AL[CHAR] |
| 462 | PColumn3 KNOWLEDGEAREA for 103(0)[232] NOT NULL |
| 495 | PColumn3 NAME for 103(1)[ CHAR] NOT NULL |
| 519 | Record 519[103] 462=447[INTEGER],495=AL/Basic Analysis[CHAR] |
| 556 | PColumn3 CATEGORY for 134(0)[157] NOT NULL |
| 585 | PColumn3 DETAIL for 134(1)[ CHAR] NOT NULL |
| 612 | PColumn3 KNOWLEDGEUNIT for 134(2)[171] NOT NULL |
| 647 | Record 647[134] 556=269[INTEGER],585= Differences among best, expected, and worst case behaviors of an algorithm[CHAR],612=519[INTEGER] |
| 750 | Record 750[134] 556=269[INTEGER],585= Asymptotic analysis of upper and expected complexity bounds[CHAR],612=519[INTEGER] |
| 838 | Record 838[134] 556=269[INTEGER],585= Big O notation: formal definition[CHAR],612=519[INTEGER] |
| 900 | Record 900[134] 556=269[INTEGER],585= Complexity classes, such as constant, logarithmic, linear, quadratic, and exponential[CHAR],612=519[INTEGER] |
| 1014 | Record 1014[134] 556=269[INTEGER],585= Empirical measurements of performance[CHAR],612=519[INTEGER] |
| 1080 | Record 1080[134] 556=269[INTEGER],585= Time and space trade-offs in algorithms[CHAR],612=519[INTEGER] |
| 1148 | Record 1148[134] 556=292[INTEGER],585= Big O notation: use[CHAR],612=519[INTEGER] |
| 1196 | Record 1196[134] 556=292[INTEGER],585= Little o, big omega and big theta notation[CHAR],612=519[INTEGER] |
| 1267 | Record 1267[134] 556=292[INTEGER],585= Recurrence relations[CHAR],612=519[INTEGER] |
| 1316 | Record 1316[134] 556=292[INTEGER],585= Analysis of iterative and recursive algorithms[CHAR],612=519[INTEGER] |
| 1391 | Record 1391[134] 556=292[INTEGER],585= Some version of a Master Theorem[CHAR],612=519[INTEGER] |
| 1452 | PColumn3 CATEGORY for 185(0)[157] NOT NULL |
| 1481 | PColumn3 DETAIL for 185(1)[ CHAR] NOT NULL |
| 1508 | PColumn3 KNOWLEDGEUNIT for 185(2)[171] NOT NULL |
| 1543 | PColumn3 LEVEL for 185(3)[218] NOT NULL |
| 1570 | Record 1570[185] 1452=269[INTEGER],1481= Explain what is meant by “best”, “expected”, and “worst” case behavior of an algorithm.[CHAR],1508=519[INTEGER],1543=359[INTEGER] |
| 1705 | Record 1705[185] 1452=269[INTEGER],1481= In the context of specific algorithms, identify the characteristics of data and/or other conditions or assumptions that lead to different behaviors.[CHAR],1508=519[INTEGER],1543=383[INTEGER] |
| 1890 | Record 1890[185] 1452=269[INTEGER],1481= Determine informally the time and space complexity of simple algorithms.[CHAR],1508=519[INTEGER],1543=406[INTEGER] |

The data captured in the above statements do not conform to ISO39075 node/edge graphs but can be simply stored in SQL tables with reference columns as described above. Then any of the formats discussed in section II above can easily be exported using JSON for input to other systems or formatted back into a readable document rather like the ACM report we began with. More importantly, it can be queried and analyzed, and further data and metadata can be added about authorship, updates, provenance, delivery by universities, and outcomes.

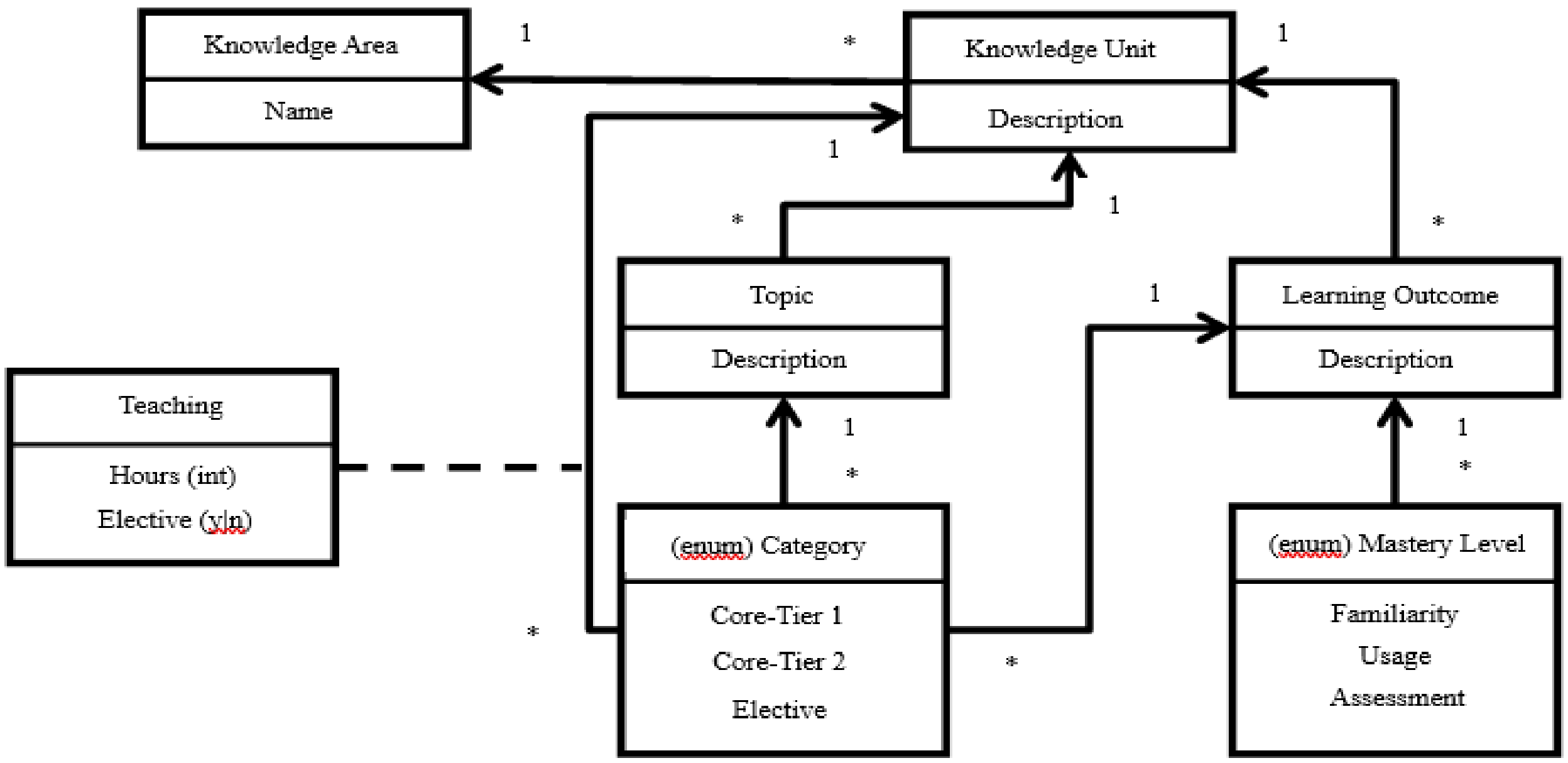

Figure 3. The implied data model

In more complex graphs, the ability to label arrows as in -AnnualFee-> is useful. Such arrow labels can be used as an alternative to labelling node types, and in combination can provide facilities for generalization and/or subtyping.

Reverse arrows can also be allowed, and in every case the node at the arrow base will contain a reference to the node at the arrowhead. With these conventions, a new label can be bound by its use on an arrow or as a node type, and on commit such bindings will be saved as a modification to the current schema (or role) in the database.

A graph schema consists of a working set of node types, edge types, arrow types and graph instances. The schema can be constructed by declaration (a closed schema), or incrementally as in the above example.

Figure 3 shows the schema implied by the above steps, with some tentative additions suggested by the text of the report [10] (a) multiplicity constraints, (b) refinement to property types (e.g., specifying the Level values as an enumeration), and (c) adding the teaching information shown in the top part of Figure 2 as a GQL-style edge.

The addition of constraints and refinement of types as in (a) and (b) here can be performed non-destructively provided the existing values satisfy the constraints or can be coerced to the new types: such schema changes in Pyrrho will modify the associated data in a cascade, whose failure will prevent any changes occurring.

## V. Implementation Highlights

In this section, we briefly review a few aspects of the implementation relevant to the above discussion of the transaction log and foreign keys.

The open-source DBMS server is v8 of PyrrhoDBMS and is available on Github. It is a record-based system with a persistent append-storage transaction log: this guarantees fully serialized isolated transactions. The transaction log is a binary file containing a sequence of objects representing a transaction start, a table definition, a column definition, a data record and so on: examples of these objects are illustrated below.

Transactions can include a mixture of data changes and schema changes. Since 2019, the implementation has used shareable data structures: these are tree structures that behave like ordinary strings to the extent that any change results in a new tree. Importantly, the server maintains these in memory as long as they are in use, and the new tree continues to use nearly all the same nodes of the old version. The only new nodes form a path from the new root to the node or modified node that has just been added. Figure 4, from [12], explains this mechanism. The result of this architecture is that every change to a database of size N takes O(logN) of processing time and storage.

This design is implemented in Pyrrho [2] using a generic tree structure `CTree<K,V>`. It is called CTree rather than BTree if both the generic parameters `K` and `V` contain values that can be compared using <,>, and =.

In SQL, a primary key declares a lookup index, while a foreign key declares a sort of reverse lookup: Pyrrho's implementation replaces the foreign key index with a hidden reference column. For example, the SQL statements

```
QL> create table company (name string primary key)
QL> create table person (name string primary key,worksin string references company)
```

result in the following entries in the transaction log:

```
23 PTable COMPANY
35 PColumn3 NAME for 23(0)[Domain CHAR]
58 PIndex COMPANY on 23(35)PrimaryKey
98 PTable PERSON
```

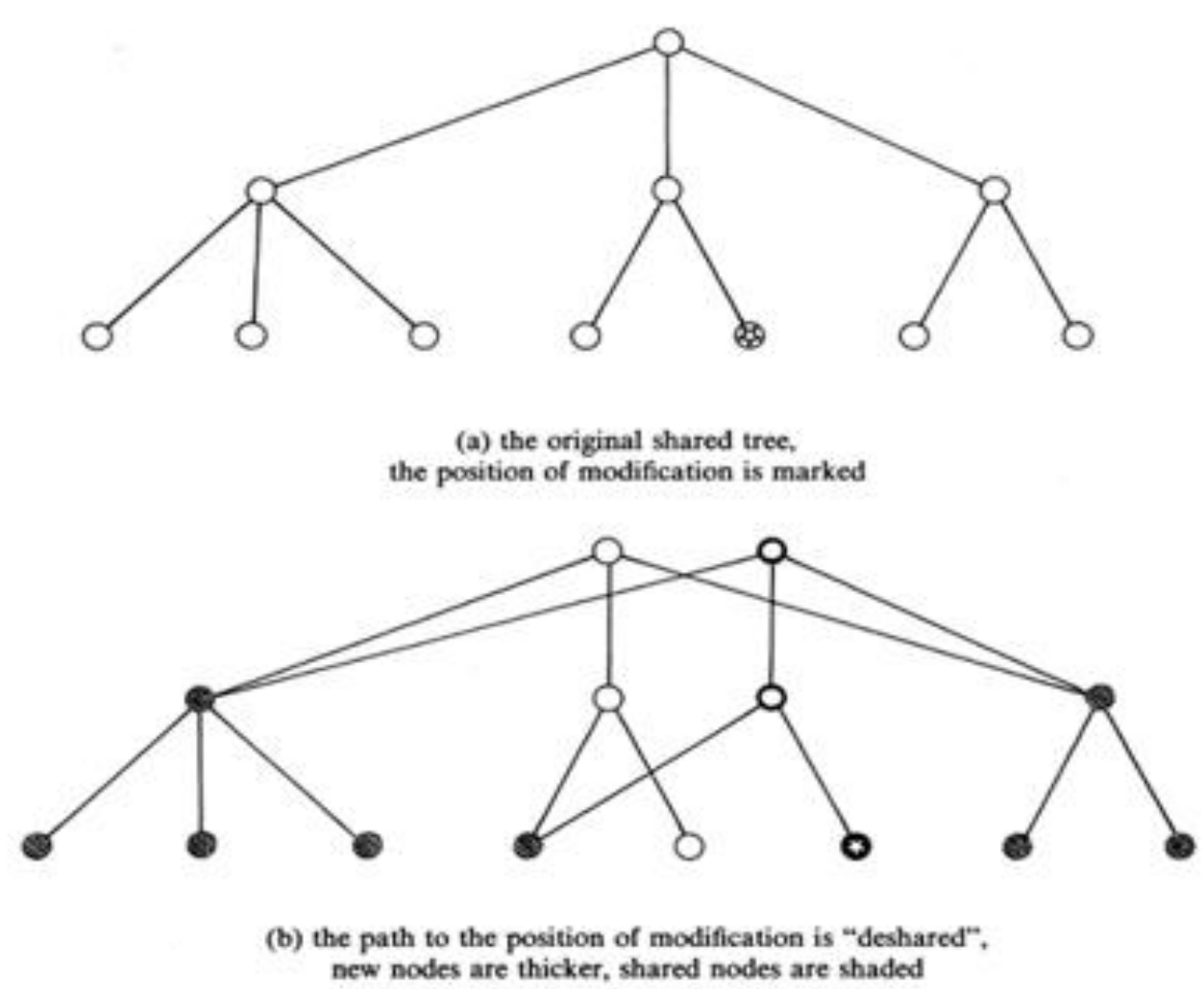

Figure 4. The shareable data concept [12].

```
109 PColumn3 NAME for 98(0)[Domain CHAR]
132 PIndex PERSON on 98(109)PrimaryKey
153 PColumn3 WORKSIN for 98(1)[Domain CHAR]
180 PIndex COMPANY on 23(153) RestrictUpdate,
RestrictDelete
204 Domain REF 23
218  PColumn3 COMPANY for 98(2)[204] RestrictUpdate,
RestrictDelete keymap 180
```

The last two entries define a reference column that is to be constructed from the value of the WORKSIN column. In the database itself, there is no object corresponding to the reverse index 180, and the Person table displays the first two columns NAME and WORKSIN:

```
{COMPANY TABLE (35) Display=1 [35,Domain CHAR]
Indexes:((35)58)}
{TableColumn NAME 35 Domain CHAR optional Table=23
RefCols: [218]}
{Index 58 for 23 Key:Domain ROW (35)
Display=1[35,Domain CHAR]
{PERSON TABLE (109,153|218) Display=2 [109,Domain
CHAR], [153,Domain CHAR], [218,Domain REF 23]
Indexes:((109)132) ColRefs: 23=[218]}
{TableColumn NAME 109 Domain CHAR optional Table=98}
{Index 132 for 98 Key:Domain ROW (109)
Display=1[109,Domain CHAR]
{TableColumn WORKSIN 153 Domain CHAR optional Table=98}
{TableColumn 218 Domain REF 23 optional Table=98 keyMap
(153) RestrictUpdate, RestrictDelete}
```

Column 218 contains values generated from the WORKSIN column 153 using the primary index of table 23. The mechanism for keeping track of references using structural row uids is detailed below. It should now be relatively clear how reference values support edge connections.

The implementation of graph types in Pyrrho is role-based and is intended to be intuitive. Given a relational database in Pyrrho and the right privileges, new graph data can be created within the current role by defining element types (under or by adding metadata to existing tables or types) or incrementally using INSERT . If the user has sufficient privileges to modify existing tables, new properties can be added explicitly or implicitly, and new edge connections implemented (with default properties FROM, TO and WITH. By default, new graph properties and references are not optional.

Existing relational data in the database can be accessed using a default graph model whereby tables without foreign keys can be accessed as node types, and tables with at least two foreign keys can be used as edge types: the first two foreign key columns become the default FROM and TO connections, and additional or replacement element types can be constructed or modified implicitly as above or by explicit create and alter statements. Importantly, in all cases the identification of graph data in the database is role-specific, sharing the underlying tables and types with other roles subject to privileges granted to them.

This approach is pragmatic rather than purist: while by abuse of language SQL types are treated internally as a subclass of SQL tables and GQL edge types a subclass of GQL node types, they occupy different namespaces and for most purposes overloading of names will not be a problem; for example, a graph pattern will search for a node label in the set of nodetypes or an edge label in the set of edgetypes, while SQL uses different syntax for tables and types. An explicit CREATE with the same name as another object of a different class will create a new object with a different uid, and any name conflicts can be resolved by role-based renaming and/or the use of numeric uids for objects hidden by name overloading. SQL tables and SQL types have different semantics for columns and fields respectively: for example, a table column may contain nulls unless declared NOT NULL while a type field is required unless declared OPTIONAL. These aspects remain true even if the table is treated as a node type, or a node type is treated as an edgetype. Importantly, SQL types support methods and UNDER, which is needed when the set of columns is extended by the the subtype ("width subtyping"), while GQL types defined on an SQL table are only able to use subtypes defined by a "depth-subtyping" constraint such as CHECK FROM IS OF(person), WHERE MATCH .., or the recemtly-proposed REQUIRE syntax.

There are some resulting extensions to SQL, allowing multiple inheritance to be permitted for graph types in addition to nesting. Pyrrho also permits n-ary edge type. A graph type can be constructed implicitly by an inline insert/create statement if it has a label without & . Pyrrho does not have a distinguished ID column: if such a column is declared it has no special status beyond being the default name for the node in visualisations of portions of the graph .

In Section IV, we saw how the arrow notation automatically adopted (or created) reference links between records. The implementation maintains a list of reference connections for each table, and these can be upgraded to edge types.

Assuming the current user has the privileges to do so (e.g., database owner), the data can be dynamically extended during data insertion: extra properties added to nodes, and single-valued properties can be converted to set-valued properties. GQL allows sets of node labels, and these simply inherit the union of properties associated with these labels. In Section III, the introduction of an edge property between, say, KnowledgeUnit and Topic, as in `-[:Syllabus{at:'Yale'}]->` would infer the creation of a table of edges specific to one or more universities, and also give a name to such edges (section III left them unnamed).

Once the data model is built, JSON can be generated directly by the server suitable for import into a large language model. Currently the server has system tables that contain these JSON items for the schema, for example table "Role$Json". In the above example, this gives the schemata shown in Tables II and III.

TABLE II. SCHEMA FOR CS2013

| Name | Definition |
|---|---|
| CATEGORY | {Fields:[NAME:{Domain:CHAR}],Display:1} |
| LEVEL | {Fields:[NAME:{Domain:CHAR}],Display:1} |
| KNOWLEDGEAREA | {Fields:[NAME:{Domain:CHAR}],Display:1} |
| KNOWLEDGEUNIT | {Fields:[KNOWLEDGEAREA:{Domain:REF,ToType:'KNOWLEDGEAREA'},NAME:{Domain:CHAR}],Display:2} |
| TOPIC | {Fields:[CATEGORY:{Domain:REF,ToType:'CATEGORY'},DETAIL:{Domain:CHAR},KNOWLEDGEUNIT:{Domain:REF,ToType:'KNOWLEDGEUNIT'}],Display:3} |
| LEARNINGOUTCOME | {Fields:[CATEGORY:{Domain:REF,ToType:'CATEGORY'},DETAIL:{Domain:CHAR},KNOWLEDGEUNIT:{Domain:REF,ToType:'KNOWLEDGEUNIT'},LEVEL:{Domain:REF,ToType:'LEVEL'}],Display:4} |

TABLE III. FOR THE AUTHOR AND BOOK EXAMPLE

| Name | Definition |
|---|---|
| AUTHOR | {Fields:[ID:{Domain:INTEGER},ANAME:{Domain:CHAR}],Display:2} |
| BOOK | {Fields:[ID:{Domain:INTEGER},AUTHID:{Domain:INTEGER}, TITLE:{Domain:CHAR},AUTHOR:{Domain:REF,ToType:'AUTHOR'}],Display:3} |

In this way, graphical database concepts are simply added to the basic SQL implementation. NodeTypes are a subclass of user defined types, and EdgeTypes are a subclass of NodeTypes, but the role manages separate namespaces for each kind of object. We recall that database objects and their contents can be renamed in roles.

Graphical data is nevertheless record-based data, and so all these classes are implemented as tables. Tables and user-defined types can acquire graphical properties by adding metadata or by the operation of the graphical INSERT statement. However, properties are always implemented by columns in tables or fields in types, and edges are implemented behind the scenes using reference values, and graphical operations and specifications must be compatible with (or, in the hands of an object owner, modify) existing database structure. By default, columns in tables represent optional properties, and columns and references in user-defined types are not optional.

Pyrrho enhances the GQL and SQL model by means of subtypes and unions, on the principle that additional properties imply subtypes. A new node type in GQL can have a set of type labels, and if these are existing types or tables in the database, all of their properties are properties of the new type. An unlabeled node in a Match statement is constrained by adjacent nodes and edges to belong to a particular table or type.

In this way, information about database objects can be role-specific and dynamic. The database objects affected have system-managed properties that reflect graphical aspects of the current database contents as follows:

SysRefIndex keeps track of reference values. It is a database-level property of any referenced Table T of form `CTree<long,CTree<long,CTree<long,bool>>>`. The first long is the uid of a referenced row rr in T, the second is the uid of a referencing column rc in another Table S, and the third is the uid of a referencing row ss in S. Then `rr = ss.vals[rc]`, so that SysRefIndex provides a reverse index to referencing data.

ColRefs is a list of referencing columns in a given Domain D, of form CTree<long,CTree<long,bool>>. The first long is a Table referenced by D, while the second is the referencing column in D.

Model is a role-dependent (or schema-dependent) property of any database object S that contains references accessed using type names or named arrows, of form `CTree<string,CTree<Qlx,long>>`. Here the string is the name of a referenced table T, the token is an arrow, and the long identifies a reference column of S. A useful convention is to give a simple reference column the name of the referenced table, and the default arrow name is the empty string so the statement parser can use a simple algorithm to compute the reference column for each reference.

The server is a multi-threaded TCP server using platform- and culture-independent strong primitive data types. It is implemented for Microsoft .NET Core 10.0 and so will run on Windows, Linux and MacOS.

## VI. PERFORMANCE MEASUREMENTS

As mentioned above, there are good reasons to believe that the performance of the engine as a DBMS will be satisfactory. PyrrhoDBMS builds databases completely in memory using disk storage merely for the persistent transaction log. For the Graphics Data Council Financial Benchmark [13], we obtained the results shown in Table IV.

TABLE IV. BUILDING THE FINANCIAL BENCHMARK

| Scale | Build time (seconds) | Ratio | Log Size (KB) | Ratio |
|---|---|---|---|---|
| SF001 | 83 | | 5768 | |
| SF01 | 868 | 10.46 | 57383 | 9.95 |
| SF1 | 9507 | 10.96 | 579151 | 10.09 |

For this experiment, the benchmark was modified as shown in the Appendix to showcase subtypes and ternary edges and use a combination of SQL and GQL syntax. The task was to build the database from the spreadsheets that use the original data model provided by LDBC. For all of the many ternary edges (subclassing `TRANSFER`) a third reference property was calculated using a default expression that uses the Match statement, highlighted in the Appendix. Performance of the DBMS shows very good scalability. It showed that the performance was nearly linear in the construction time for graphical models with transaction log size 5 MB, 50 MB, 500 MB. The amount of memory used to hold the database was also nearly linear, at roughly 10 times the transaction log size. Timings for a more general workload will depend on the complexity of the workload and (for example) whether multiplicity constraints require to be rechecked on every data modifying transaction.

## VII. CONCLUSIONS AND FUTURE WORK

This work is primarily a continuation of our integration of graphical database ideas into our SQL implementation, and the provision of support for resource description using JSON. It seems clear that the use of references to data entities as records is simpler than using lists of key values, especially when navigating links between records. For this reason, we believe that our work will support greater integration of real-world data into systems that seek to build models of expected behavior, such as resource augmented generation.

## APPENDIX

```
create type legalentity as (name string, isBlocked
boolean, createtime timestamp, country string, city
string) nodetype
create type person under legalentity as (id int primary
key, gender string, birthday timestamp)
create type company under legalentity as (cid int
primary key, business string, description string,url
string)
create type actbase as (createtime timestamp, accttype
string) nodetype
alter table actbase add balance float64 default 0.0
create type account under actbase as (id int primary
key, isBlocked boolean, nickname string, phonenumber
string, email string, freqlogintype string,
lastlogintime timestamp, accntlevel string)
create type loan under actbase as (lid int primary key,
loanAmt float64, interest float32)
create type medium as (id int primary key, type string,
isblocked boolean, createtime timestamp, lastlogin
timestamp, risklevel string) nodetype
create type apply as (createTime timestamp,
organization string) edgetype (legalentity, loan)
create type personapplyloan under apply edgetype
(person, loan)
create type companyapplyloan under apply edgetype
(company, loan)
create type guarantee as (createTime timestamp,
relationship string) edgetype (legalentity,legalentity)
create type personguaranteeperson under guarantee
edgetype (person, person)
create type invest as (ratio float64, createTime
timestamp) edgetype (legalentity, legalentity)
create type personinvestcompany under invest edgetype
(person, company)
create type owns as (createTime timestamp) edgetype
(legalentity, actbase)
create type personownsaccount under owns edgetype
(person, account)
create type activatedfor as (createtime timestamp,
location string) edgetype (med=medium, account)
create type transfer as (amount float64, createTime
timestamp, orderNumber int optional, comment string
optional, payType string optional, goodsType string
optional) edgetype (from actbase with activatedfor
optional to actbase)
alter table transfer alter with1 to usedFor
alter table transfer alter usedFor add default (match
any 1 (from1)<-[a:activatedfor where
a.createtime>=from1.createtime]-() yield a)
create type accountrepayloan under transfer edgetype
(from account to loan)
create type accounttransferaccount under transfer
edgetype (from account to account)
create type accountwithdrawaccount under transfer
edgetype (from account to account)
create type companyguaranteecompany under guarantee
edgetype (company, company)
create type companyinvestcompany under invest edgetype
(company, company)
create type companyownsaccount under owns edgetype
(company, account)
create type loandepositaccount under transfer edgetype
(from loan with activatedfor optional to account)
```